\documentclass{aastex63}
\usepackage{upgreek}
\usepackage{amsmath}
\usepackage[T1]{fontenc} %% For some names in references with \k{E}

\submitjournal{ApJ}

\shorttitle{Soft X-ray Spectral Diagnostics of Solar Flares}
\shortauthors{Mithun, N. P. S. et al.}

\begin{document}

\title{Soft X-ray Spectral Diagnostics of Multi-thermal Plasma in Solar Flares 
with Chandrayaan-2 XSM}

\correspondingauthor{N. P. S. Mithun}
\email{mithun@prl.res.in}

\author[0000-0003-3431-6110]{N. P. S. Mithun}
\affiliation{Physical Research Laboratory, Navrangpura, Ahmedabad, Gujarat-380 009, India }
\affiliation{Indian Institute of Technology Gandhinagar, Palaj, Gandhinagar, Gujarat-382 355, India}
\author[0000-0002-2050-0913]{Santosh V. Vadawale}
\affiliation{Physical Research Laboratory, Navrangpura, Ahmedabad, Gujarat-380 009, India }
\author[0000-0002-4125-0204]{Giulio Del Zanna}
\affiliation{DAMTP, Centre for Mathematical Sciences, University of Cambridge, Wilberforce Road, Cambridge, CB3 0WA, UK}
\author[0000-0002-8050-924X]{Yamini K. Rao}
\affiliation{DAMTP, Centre for Mathematical Sciences, University of Cambridge, Wilberforce Road, Cambridge, CB3 0WA, UK}
\author[0000-0001-5042-2170]{Bhuwan Joshi}
\affiliation{Physical Research Laboratory, Navrangpura, Ahmedabad, Gujarat-380 009, India }
\author[0000-0002-4781-5798]{Aveek Sarkar}
\affiliation{Physical Research Laboratory, Navrangpura, Ahmedabad, Gujarat-380 009, India }
\author[0000-0002-7020-2826]{Biswajit Mondal}
\affiliation{Physical Research Laboratory, Navrangpura, Ahmedabad, Gujarat-380 009, India }
\author[0000-0003-2504-2576]{P. Janardhan}
\affiliation{Physical Research Laboratory, Navrangpura, Ahmedabad, Gujarat-380 009, India }
\author[0000-0003-1693-453X]{Anil Bhardwaj}
\affiliation{Physical Research Laboratory, Navrangpura, Ahmedabad, Gujarat-380 009, India }
\author[0000-0002-6418-7914]{Helen E. Mason}
\affiliation{DAMTP, Centre for Mathematical Sciences, University of Cambridge, Wilberforce Road, Cambridge, CB3 0WA, UK}

\begin{abstract}
Spectroscopic observations in X-ray wavelengths provide excellent diagnostics of the temperature distribution 
in solar flare plasma.
The Solar X-ray Monitor (XSM) onboard the Chandrayaan-2 mission provides broad-band disk integrated soft X-ray 
solar spectral measurements in the energy range of 1-15 keV with high spectral resolution and time cadence.
In this study, we analyse X-ray spectra of three representative GOES C-class flares obtained 
with the XSM to investigate   the   evolution of  various   plasma parameters during the course of the flares.
Using the soft X-ray spectra consisting of the continuum and well-resolved line complexes of major elements like 
Mg, Si, and Fe, we investigate the validity of the isothermal and multi-thermal assumptions on the high temperature 
components of the flaring plasma. We show that the soft X-ray spectra during the impulsive phase of the 
high intensity flares are inconsistent with isothermal models and are best fitted with double peaked 
differential emission measure distributions where  the temperature of the hotter component 
rises faster than that of the cooler component. The two distinct temperature components observed in DEM models 
during the impulsive phase of the flares suggest the presence of the directly heated plasma in the corona and  
evaporated plasma from the chromospheric footpoints.
We also find that the abundances of low FIP elements Mg, Si, and Fe reduces from near coronal to near photospheric values 
during the rising phase of the flare and recovers back to coronal values during decay phase, which is also consistent with the 
chromospheric evaporation scenario.
\end{abstract}

\keywords{Sun: X-rays  -- Sun: corona -- Sun: Flares}

\section{Introduction} 
\label{sec:intro}

Solar flares are sudden releases of energy in the lower atmosphere of the Sun. They emit 
across the entire electromagnetic spectrum and are responsible for accelerating particles. 
Various observations point to magnetic reconnection as the underlying
mechanism powering the flares, which accelerates the particles to high energies and heats the
plasma to temperatures higher than a few million kelvin~\citep{2017LRSP...14....2B}. The standard flare model,
also known as CSHKP model~\citep{1964NASSP..50..451C,1966Natur.211..695S,1974SoPh...34..323H,1976SoPh...50...85K},
has been successful in explaining several observed features of solar flares.
However, multiple aspects such as the exact location of particle acceleration and heating and the acceleration 
mechanism still remain unspecified~\citep{2017LRSP...14....2B,2008A&ARv..16..155K}.
Resolving these issues requires knowledge of the local plasma’s thermal and non-thermal particle distributions, which
can be best obtained from the X-ray and EUV observations~\citep{2018LRSP...15....5D}.

More often than not, flare thermal plasma is composed of particles at multiple temperatures.
The differential emission measure (DEM) is a way to quantify the thermal plasma at different temperatures and densities 
along the line-of-sight. 
Ideal observations to required obtain the DEM in flaring regions would be spatially resolved
measurements of individual spectral lines of various elements
formed at different temperatures. While there are concept designs of instruments
that may be capable of such measurements~\citep{2010arXiv1011.4052L,2021ExA...tmp..135M,2021FrASS...8...33D},
there are currently no such instruments available.

However, it is possible to derive spatially resolved DEMs by using the multi-band
images in EUV with the SDO AIA~\citep{2015ApJ...807..143C,2018ApJ...856L..17S} or in X-rays
with the Hinode XRT~\citep{2010A&A...523A..44G}, both of which provide  limited spectral information.
In the case of AIA, even though the wavelength bands are relatively narrow, they include multiple spectral lines
that are formed at different temperatures. In the case of XRT, in addition to lines formed at different
temperatures, there is a significant contribution from the continuum which is strongly dependant on
temperature. This results in broad temperature responses for different wavelength bands
making it difficult to obtain unique DEM solutions.

Another approach is to use high resolution spectra in the EUV or X-rays, but without any spatial information,
where individual lines formed at different temperatures are well resolved.
Earlier observations with SMM/BCS, Yokoh/BCS, and other crystal spectrographs in X-ray wavelengths have 
provided evidence of multi-thermal plasma distributions~\citep{1990ApJS...73..117D,1996PhPl....3.3203F}.
More recently, DEM distributions have been obtained with this
approach using observations with SDO EVE~\citep{2013ApJ...770..116W} and RESIK~\citep{2014ApJ...787..122S,2018JASTP.179..545K} observations.
Observations in X-rays are more sensitive to hotter plasma ($\sim>10$ MK) than EUV observations and 
thus combining both EUV and higher energy X-ray 
observations can be advantageous.
For example, \cite{2014ApJ...788L..31C} and \cite{2019ApJ...881..161M} have carried out joint analyses of EUV spectra from
EVE along with X-ray observations ($> 3$ keV) with RHESSI to obtain better constraints on the DEM. Similar studies
have also been carried out recently with the FOXSI hard X-ray imaging telescope~\citep{2020ApJ...891...78A}.
As observations in lower energy X-rays down to 1 keV complement the temperature ranges covered by EUV and X-ray 
observations at higher energies by RHESSI,
there have also been attempts to use spatially and spectrally integrated soft X-ray measurements
from GOES XRS along with EUV
spectra to constrain the DEM~\citep{2014ApJ...786L...2W}.
The broadband counts from GOES, dominated by the continuum~\citep{2013A&A...555A..59D}, provide limited temperature diagnostics in 
comparison to having the complete spectral data.

Spatially integrated, but spectrally resolved soft X-ray observations can provide better constraints 
to the DEM and complement EUV and hard X-ray observations.There have been a few instruments in the past such
as SOXS(4 -- 25 keV,~\citealp{2005SoPh..227...89J}, SphinX(1 -- 15 keV, ~\citealp{2013SoPh..283..631G}), and MinXSS(1 --15 keV, 
~\citealp{moore18}) 
that have provided soft X-ray spectral measurements.
Often soft X-ray spectra from these instruments are fitted with isothermal models; however, there have also been studies that have used 
X-ray spectral data to obtain the differential emission measure~\citep{2016ApJ...823..126A}.

If the soft X-ray spectra have sufficient energy resolution to resolve the line complexes
of individual elements, then it is also possible to constrain and measure the abundances of elements 
during the course of flares, self consistently with DEM. Recent reports of measurements of 
elemental abundances during flares obtained by isothermal fits to the soft X-ray spectra 
show close to photospheric abundances during the flare peaks~\citep{2021ApJ...920....4M, 2020SoPh..295..175N}.

Here, we attempt to investigate the evolution of the temperature structure and elemental abundances  
during solar flares using soft X-ray observations with the Solar X-ray Monitor (XSM) on board the Chandrayaan-2 
mission~\citep{2014AdSpR..54.2021V,shanmugam20,2020SoPh..295..139M}.
The XSM provides disk-integrated spectral observations in the energy range of
1 -- 15 keV with the highest energy resolution thus far for such broadband spectrometers.
Soft X-ray spectra obtained with the XSM for sub-A class events
have been shown to be consistent with an isothermal model~\citep{2021ApJ...912L..13V}.
It is also observed that the spectra are well described with a single temperature plasma
emission model during the evolution of B-class flares~\citep{2021ApJ...920....4M}.
However, for more complex and larger flares such as C-class events, we explore here the
validity of isothermal models in fitting the observed spectrum. We then propose a scheme for
DEM analyses of X-ray spectra, such as that with the XSM, assuming simple functional forms for
the DEM distribution. These methods are employed to obtain the evolution of the DEM during 
the course of the flare along with the evolution of abundances of some of the elements.

Section~\ref{sec:obs} provides details of the observations of the flares considered in the present work
and the data reduction. Analysis of the spectra with an isothermal approximation is presented in Section~\ref{sec:isoth} and
Section~\ref{sec:DEM} presents the analysis and results considering a multi-thermal plasma.
Results are discussed and summarized in Section~\ref{sec:discussion}.

\section{Observations of Flares with XSM and Data Reduction}
\label{sec:obs}

The Chandrayaan-2 Solar X-ray Monitor (XSM) carries out Sun-as-a-star observations and measurements of the 
1--15 keV X-ray spectra with an energy resolution of $\sim$ 175 eV at 5.9 keV at one second 
cadence~\citep{2014AdSpR..54.2021V,shanmugam20,2020SoPh..295..139M}. 
The XSM has been observing the Sun since September 2019 and the solar X-ray light 
curves are available on the XSM website\footnote{\url{https://www.prl.res.in/ch2xsm/}}.
As the visibility of the Sun for XSM varies with orbital seasons~\citep{mithun20_gcal}, 
there are long periods without solar observations. Also, with XSM being in a lunar 
orbit, there are periods when the Sun is occulted by the Moon, thereby resulting 
in partial observations of solar flares. 
For the present work, we selected three representative GOES C-class flares having 1-8~\AA~ peak fluxes in 
ranging from 1.5$\times 10^{-5}~\mathrm{Wm^{-2}}$ to 8.5$\times 10^{-5}~\mathrm{Wm^{-2}}$, 
for which XSM observations are available for the entire flare duration. 
The selected flares are SOL-2020-10-16T12:57 (C1.57), SOL-2021-10-07T02:46 (C5.70), and 
SOL-2021-09-08T17:32 (C8.40)      

\begin{figure}[h!]
\begin{center}
    \includegraphics[width=0.6\columnwidth]{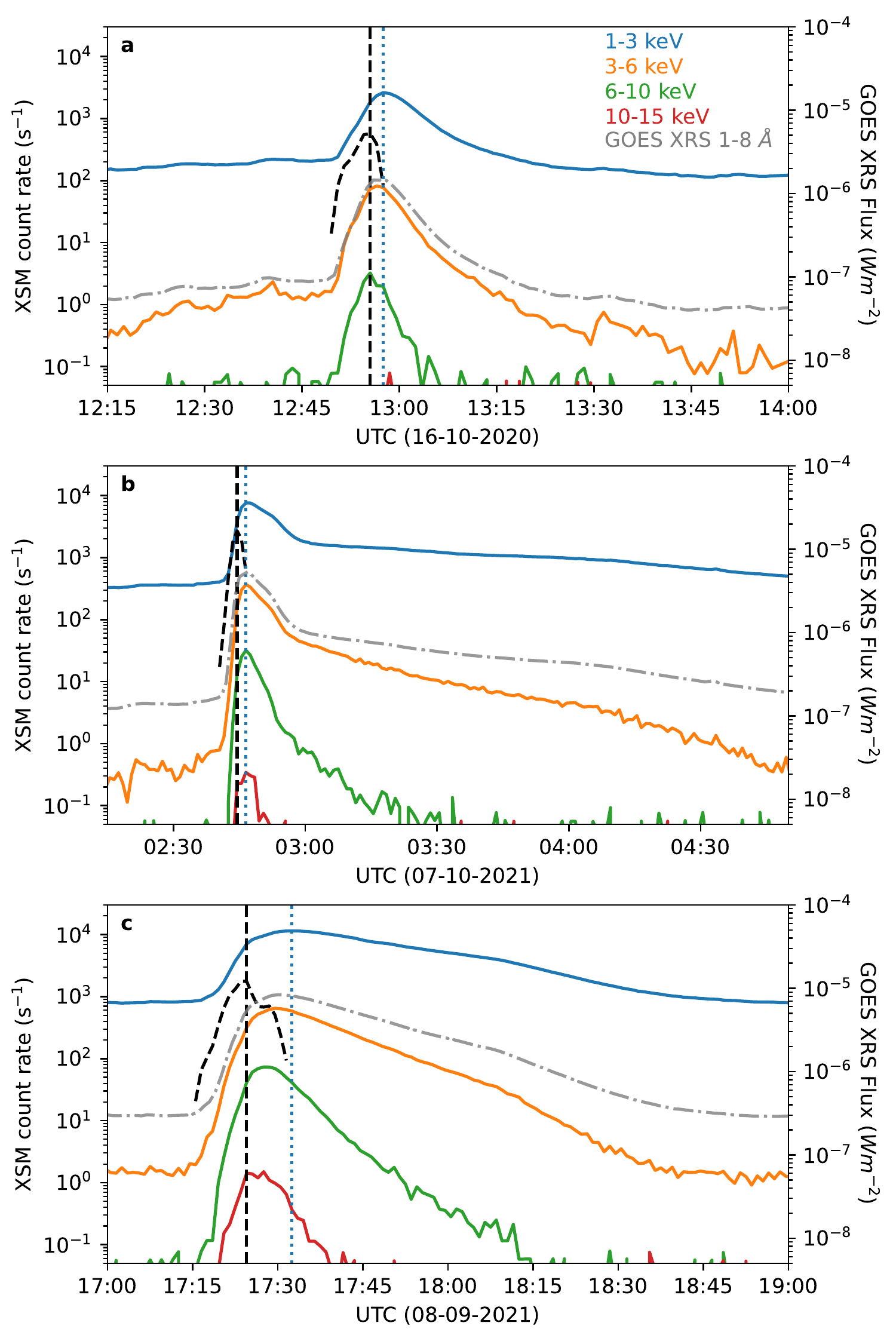}
    \caption{XSM light curves in different energy bands for the three C-class 
    flares on 16-Oct-2020 ({\bf a}), 07-Oct-2021 ({\bf b}), and 08-Sep-2021 ({\bf c}). GOES XRS 1-8~\AA 
    ~band flux is shown in grey. All light curves are at one minute cadence. 
    The y-range is selected such that the points shown have at least 2 sigma detection.
    The derivative of 
    the XSM 1--15 keV count rate is shown with a black dashed line and the vertical black dashed 
    line denotes the peak of the impulsive phase where the derivative is maximum. 
    The blue vertical dotted line corresponds to the peak of the 1--15 keV light curve.
    \label{flareLc}}
\end{center}
\end{figure}

For each of the flares, we generated X-ray light curves from XSM raw data using the \verb|xsmgenlc| 
task from the XSM Data Analysis Software~\citep{mithun20_soft} with the default 
good time intervals.
The light curves were generated  in different energy bands at a cadence of 1 minute, 
after correcting for the instrument effective area variations 
with Sun angle over time. Figure~\ref{flareLc} shows the XSM light curves in different 
energy bands for the three flares. For comparison, X-ray flux light curves from 
the GOES-16/17 XRS instrument in the 1--8 \AA ~band are also shown in the 
figure. 

It can be seen that all three flares follow the general characteristics , i.e. an impulsive phase and a gradual phase which lasts progressively longer for larger flares. 
Light curves in different energy bands follow the 
expected nature that the higher energy bands peak earlier  compared to progressively 
lower energy bands.
We also compute the derivative of the soft X-ray (1--15 keV) XSM light curve which is 
expected to be similar to the impulsive hard X-ray emission by the Neupert effect~\citep{1968ApJ...153L..59N}. We denote 
the peak of the flux derivative as the impulsive phase peak, marked by vertical 
dashed lines in the figure. Times corresponding to the peak count rate in 1--15 keV 
band are also shown in the figure with vertical dotted lines. 
 
Further, to carry out a spectral analysis with XSM data, time resolved spectra are generated for one minute intervals
using \verb|xsmgenspec|. A one minute time bin was chosen to ensure that a sufficient number 
of counts are present in the spectra to carry out spectral fitting. For each spectrum, a corresponding 
ancillary response file with effective area was also generated. The spectra and response 
files are in standard formats that are compatible with the XSPEC spectral fitting package~\citep{arnaud96} 
used for modeling the spectra.  
The default bin size for spectral channels is 33 eV. At higher energies, the counts 
in individual channels are low, so such channels were grouped together using the \verb|grppha|
task available as part of the HEASOFT distribution. 
For each spectrum, the 1.0-1.3 keV band where the instrument
effective area is not well modelled~\citep{2020SoPh..295..139M} 
is ignored in the analysis and the spectrum above 1.3 keV is fitted. 
Using the XSM observations when the Sun
was out of its field of view, the non-solar background spectrum was generated and subtracted from the 
observed total spectrum during the flares. 
Solar spectra have significantly higher counts compared to the background at low energies, but 
not at high energies.
Thus, at higher energies, spectral fitting is limited to the energy range
where the observed counts are higher than the background counts. It may be noted 
that the energy range differs for each spectra depending on the solar flux.

\section{ 
Isothermal Approximation	
}
\label{sec:isoth}

Solar flare spectra in the soft X-ray band are often approximated with isothermal 
emission models with temperature, volume emission measure, and abundances of various species 
as parameters.
It has been shown that the soft X-ray spectra of sub-A class flares observed by the 
XSM can be well described using isothermal models~\citep{2021ApJ...912L..13V}.
Further, for all B-class 
events observed during the first two seasons, 
\cite{2021ApJ...920....4M} have shown that the time-resolved spectra during the 
course of the flare are also consistent with isothermal models. 
Being the simplest model, we first analyze the spectra of the C-class flares considered for the present study 
using an isothermal approximation.

We use PyXSPEC\footnote{\url{https://heasarc.gsfc.nasa.gov/xanadu/xspec/python/html/index.html}}, 
the python interface for XSPEC, for analysis of the time-resolved spectra 
generated for one-minute intervals as discussed in the previous section.
The spectra are fitted with a model for isothermal emission computed using the CHIANTI atomic 
database version 10.0~\citep{2021ApJ...909...38D}.
The spectral model \verb|chisoth|\footnote{\url{https://github.com/xastprl/chspec}}
is implemented as a local model in XSPEC and uses tabulated 
spectra computed using CHIANTI for each elemental species (up to Zn) over 
a grid of temperatures to obtain model spectra for any temperature, emission measure, 
and abundances, as described in \citet{2021ApJ...920....4M}.
While fitting the flare spectra, abundances of all elements are initially set to 
coronal abundances from~\cite{1992PhyS...46..202F}. Then, for each spectrum,  
major elements whose line complexes are within the energy range considered for fitting 
are identfitied. 
Abundances of those elements are left as free parameters in the fitting process along with 
temperature and emission measure.
For example, if the detected spectrum extends beyond the Fe line complex at $\sim 6.5$ keV, 
the abundance of Fe is considered as a free parameter.
As the Mg and Si lines are well detected in spectra for all one minute intervals, 
their abundances are left as free parameters in the fitting of all spectra. 
Abundances of S, Ar, Ca, and Fe are allowed to vary for some of the intervals 
depending on the detection of the corresponding line complex in the spectrum.
We also note that while the XSM observations include the emission from quiescent active 
region as well, it is observed that even for B-class events the effect of this on the inferred 
parameters of the flare emission is negligible~\citep{2021ApJ...920....4M}. Hence we do not 
include pre-flare quiescent emission component separately in this work.  
  
\begin{figure}
\begin{center}
    \includegraphics[width=0.99\textwidth]{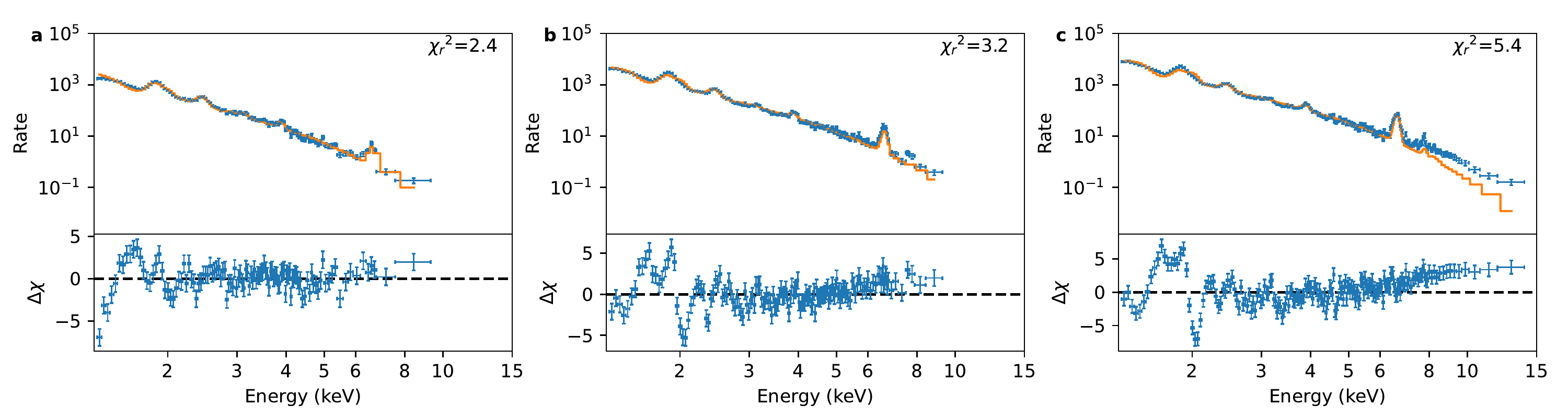}
	\caption{XSM spectra of the three flares (panel {\bf a}: C1.57, panel {\bf b}: C5.70, panel {\bf c}: C8.40) at their impulsive phase peak (marked by vertical dashed 
    lines in Figure~\ref{flareLc}) fitted with an isothermal 
    model. Orange lines show the best fit models and the lower panels show the residuals. 
    \label{isothFit}}
\end{center}
\end{figure}

Figure~\ref{isothFit} shows the observed spectra for a one minute interval at the peak of 
impulsive phase of the three flares. Best fit isothermal models are overplotted and 
the residuals are shown in the lower panels. For the weakest event in the sample (2020-10-16), 
the isothermal model fits the spectra reasonably well, although there are still 
some systematic residuals present. However, for the other two events, the residual 
of the spectra with respect to the model and the goodness of fit given in terms 
of the reduced chi-squared suggest that a single temperature 
isothermal model does not describe the observations very well.  

\begin{figure}
\begin{center}
    \includegraphics[width=0.99\textwidth]{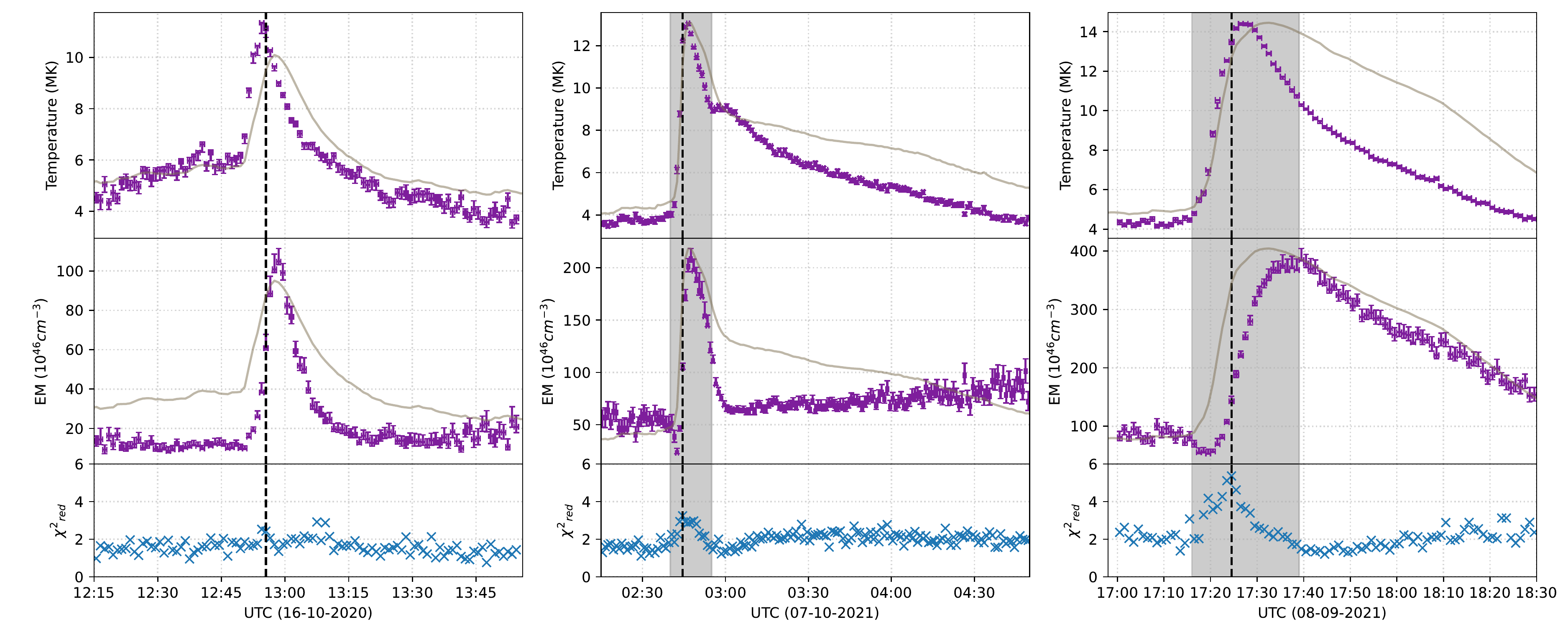}
    \caption{Results of an isothermal fit to the time-resolved spectra of the three flares. 
    Best fit temperature and emission measure with one sigma uncertainties are shown 
    in the top and middle panels. Lower panels show the reduced chi-squared of the 
    spectral fits. Impulsive phase peak times are marked by the vertical dashed lines. 
    For the two brighter events, the reduced chi-squared is seen to be higher around 
    the impulsive peak, marked by the grey shaded region.
    \label{isothTEM}}
\end{center}
\end{figure}

A similar analysis is carried out for each one minute interval during the flares and
the best fit isothermal temperature and emission measure along with the reduced 
chi-square for the entire duration of the flares are shown in Figure~\ref{isothTEM}.
It can be seen from the figure that the isothermal approximation is valid during the 
pre-flare and decay phases for all three events where the reduced chi-square are 
within acceptable ranges. For the flare on 16-October-2020, the fit is reasonable for 
the entire duration of the event. However, during the impulsive phase of the other two 
events, marked by the grey shaded region in the figure,
the reduced chi-square values suggest that the isothermal approximation 
is no longer valid. The vertical dashed lines in the figure correspond to 
the peak of the impulsive phase as identified in Figure~\ref{flareLc} and it can 
be seen that the departure from the isothermal model is maximum around these 
times. 

We also note that during the period where the observed spectra deviate from an isothermal model, 
the fitted temperature and emission measure show rapid variations. From the analysis presented 
in Appendix~\ref{sec:tavg}, we conclude that the fast evolution of temperature and emission 
measure cannot explain the deviations from isothermal models and thus the 
XSM spectra during the impulsive phase
show the presence of multi-thermal plasma.

\begin{figure}[h!]
\begin{center}
    \includegraphics[width=0.65\columnwidth]{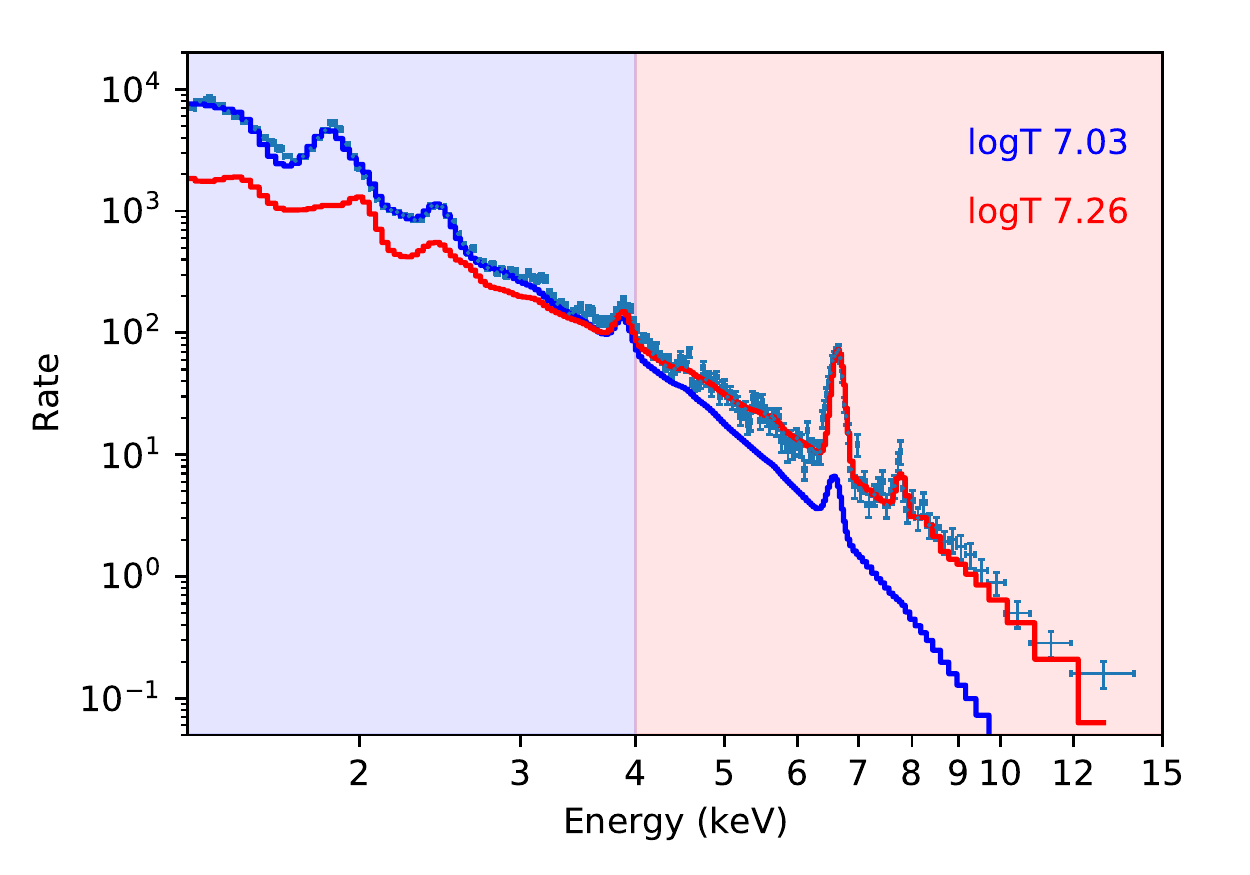}
    \caption{Spectrum during the impulsive phase of the 08-Sep-2021 flare (17:24-17:25 UTC) fitted with two different isothermal 
models considering only part of the spectrum. The spectrum below 4 keV (blue shaded interval) is fitted to 
obtain the best fit model shown in blue and a fit to the spectrum above 4 keV (red shaded interval) 
gives the best fit model shown in red.
    \label{twoTintwoEne}}
\end{center}
\end{figure}

In order to obtain an idea of the range of temperatures involved, we separately fitted the low energy part of the 
spectrum that is more sensitive to lower temperature and the high energy part of the spectrum that is more sensitive 
to higher temperatures. Spectrum below 4 keV and spectrum above 4 keV were fitted with isothermal models as 
shown in Figure~\ref{twoTintwoEne}. It can be seen that a model with logT 7.03 provides a reasonable fit to 
the low energy part of the spectrum while severely under predicting the emission at higher energies. 
Conversely, a model corresponding to logT of 7.26 is sufficient to explain the high energy part of the 
spectrum including the Fe line complex, but significantly deviates from the low energy spectrum and also 
predicts very different line shapes for the Mg/Si/S line complexes compared to the observations. 
It should also be noted that the abundances of elements were frozen to the best fit values obtained from an isothermal 
analysis. However, we find that any reasonable variations in the abundances within the ranges from photospheric 
to coronal values cannot account for the differences in line strengths between the observation and the model.
From this we conclude that at the impulsive peak, temperature distributions extending at least over this range 
of two temperatures is required to consistently explain the observed spectrum and 
we explore this in detail in the next section.

\section{Multi-thermal Plasma: Differential Emission Measure Analysis}
\label{sec:DEM}

The analysis in the previous section suggests that multi-thermal plasma is present  
during the impulsive phase of flares. The distribution of plasma at different temperatures 
can be described by the Differential Emission Measure (DEM), which provides the 
amount of plasma along the line of sight having temperature between T and T+dT~\citep{2018LRSP...15....5D}. 
With non-imaging full disk observations, one cannot obtain the spatial distribution 
of the plasma and in such cases we measure the differential on the volume emission measure (cm$^{-3}$) 
over temperature (T) defined as
\begin{equation}
	DEM (T) = \frac{d ({N_{\mathrm{e}}}^2 V)}{dT} (\mathrm{{cm}^{-3}K^{-1}})
\end{equation}  
where $V$ is the volume and ${N_{\mathrm{e}}}$ is the electron density. Conversely, the volume emission measure 
is the integral of DEM over temperature as:
\begin{equation}
	EM =\int{DEM(T)}~dT
\end{equation}

Several methods have been developed to infer the DEM from observations in different wavelengths. 
Two commonly used approaches are direct inversion and forward fitting by ${\chi}^2$ minimization (see \citet{2018LRSP...15....5D} 
for a review of methods for DEM estimation). We follow the ${\chi}^2$ minimization approach 
for estimating DEM from the XSM spectra. Often, counts over broad energy bins obtained 
from the spectra are used as input in DEM analysis. However, based on the analysis of 
temperature responses of the XSM presented in Appendix~\ref{sec:lineprof}, we find that using the full 
spectrum as input to DEM analysis provides better diagnostic potential. 
Thus, the observed spectrum as used in the isothermal analysis previously is 
used for DEM fitting.     

\subsection{Emission measure loci}

A simple method to assess the distribution of plasma at different temperatures 
without DEM fitting
is to follow the emission measure loci approach~\citep{2018LRSP...15....5D}. 
In this approach, the ratio of the observed counts in different spectral channels to the 
expected counts in the respective channels at different temperatures obtained from 
theoretical calculations are plotted. The loci 
of these curves provides an upper limit to the emission measure distribution. 
If the entire spectrum is consistent with isothermal emission, the curves of different 
channels intersect at the same point.  	

\begin{figure}[h!]
\begin{center}
    \includegraphics[width=0.65\columnwidth]{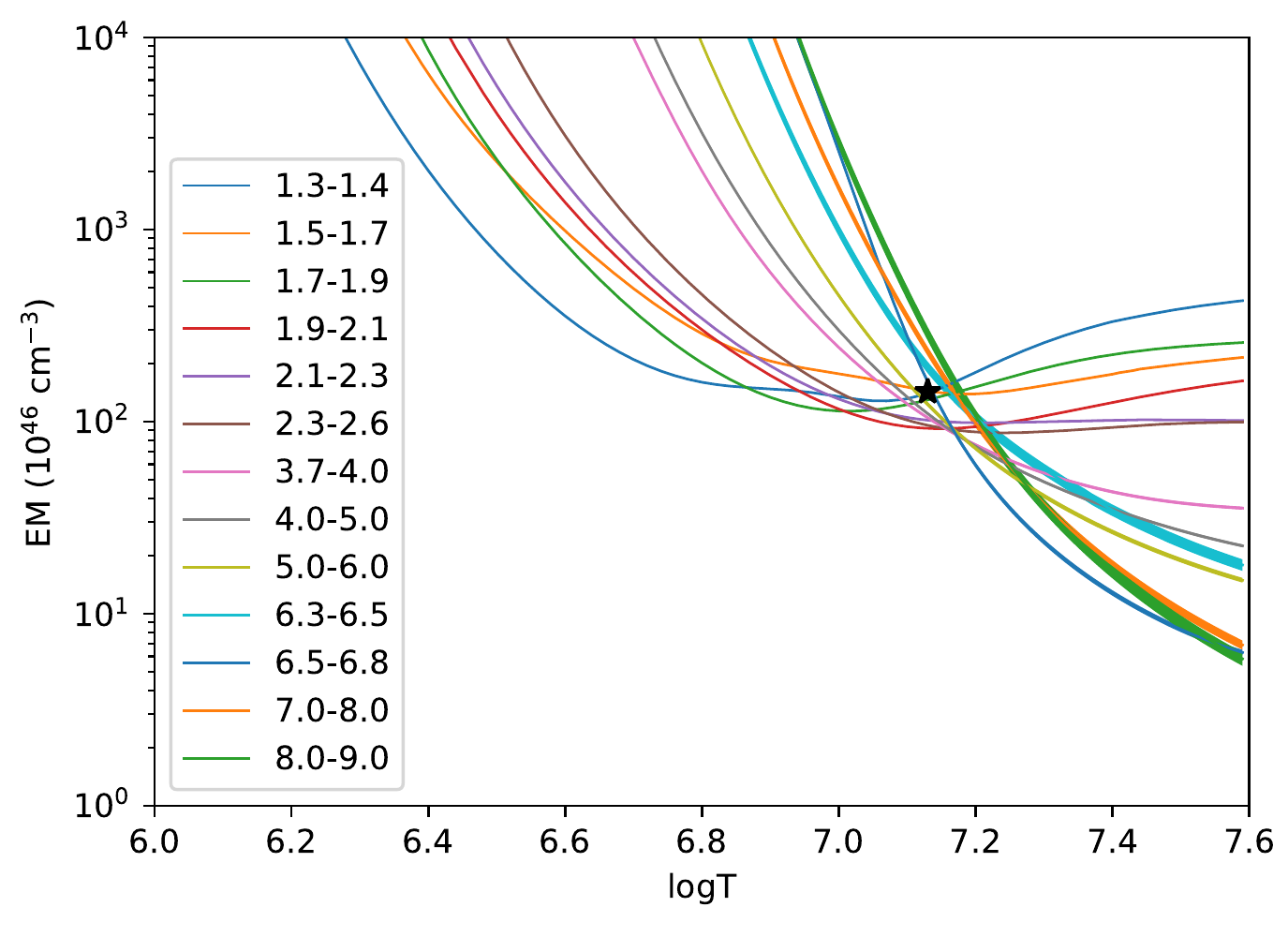}
    \caption{EM loci curves obtained from XSM spectrum during the impulsive peak of the 08-Sep-2021 flare. 
Each line corresponds to the ratio of the observed counts in the given energy range 
to the counts predicted by model for an isothermal plasma with the EM of 1 cm$^{-3}$ at different temperatures.
The width of the lines takes into account the one sigma 
uncertainties of the observed counts.  
The best fit isothermal temperature and emission measure is shown by the black star.
    \label{emloci}}
\end{center}
\end{figure}

We plot the EM loci curves for the XSM spectrum at the impulsive peak of the 08-Sep-2022 flare 
in Figure~\ref{emloci}. 
While  in principle there can be a curve for each spectral channel of XSM, in order to avoid over-crowding of 
the figure, we show loci curves for spectral channels
near the different line complexes and of some continuum channels grouped together. 
The expected counts in each channel range are obtained from the model spectra convolved with 
the instrument spectral response as was done in obtaining the temperature response discussed in Appendix~\ref{sec:lineprof}.
It can be seen from the figure that the curves do not intersect at a single point. While the isothermal 
fit result shown by the black star in the figure is at the intersection of a few loci curves, many others 
do not pass through that point as one would expect based on the residuals seen in the isothermal fits.

\subsection{Multi-thermal model fitting}

EM loci curves above provide an upper limit to the true 
DEM distribution. 
While the DEM distribution can have any shape below this upper limit, in order to obtain some 
estimates of the DEM, we start with few simple functional forms.
We note that for the DEM estimation, it is necessary to use all spectral channels of the detector, 
as discussed in the Appendix~\ref{sec:lineprof}. 
Thus, we follow a forward folding approach to estimate the DEM distribution from the XSM spectra.
Model spectrum from an assumed DEM distribution defined by few parameters is computed and then fitted to the 
observation after folding through the instrument response using PyXSPEC to obtain the best fit DEM parameters. 
We consider three multi-thermal models in the present analysis viz. two-temperature, 
Gaussian DEM, and double Gaussian DEM.

The simplest extension of an isothermal model is a two temperature model which is the 
sum of two isothermal models with independent temperatures and emission measures. 
This DEM with two delta functions is considered as the analysis in the previous section 
suggested that two different temperatures could explain the lower and higher energy 
part of the spectrum. Similar two-temperature models have been used in previous studies to 
model flare spectra(for e.g., \citealp{2010ApJ...725L.161C}).
The observed spectra are fitted with a model defined as the addition of two isothermal models 
(described in the previous section) with two temperatures, respective emission measures, and abundances 
of different elements as free parameters. 

For a broader temperature distribution, a natural choice is a Gaussian distribution\citep{2015SoPh..290.2733A} 
defined by the temperature at the peak EM ($T_p$), the peak emission measure (${EM}_p$), 
and the width of the Gaussian ($\sigma$) as: 

\begin{equation}
DEM (T) = {EM}_p exp\Bigg(-\frac{{(log(T)-log(T_p))}^2}{2{\sigma}^2} \Bigg)
\end{equation}

We have implemented a model providing the X-ray spectra for Gaussian DEM within XSPEC, 
named \verb|chgausdem|\footnote{\url{https://github.com/xastprl/chspec}}, 
for the analysis of the data. This is obtained by the weighted addition 
of spectra over the grid of temperatures with the emission measure at each temperature 
as defined by the Gaussian function. Much like the isothermal model, the Gaussian DEM 
model also has provision to vary abundances of individual elements.  

While a single Gaussian allows a broad DEM, it is constrained to have only a single maximum. 
To accommodate multiple peaks in the DEM, it can be approximated with multiple Gaussians\citep{2014ApJ...788L..31C}. However, 
the number of free parameters will be prohibitively large in that case. Instead we consider a
DEM function composed of two Gaussians with their peak temperatures, emission measures, and widths 
as independent parameters. This model is a more realistic version of the two-temperature isothermal 
model allowing finite width to the EM distributions at two temperatures. 

\subsubsection{DEM fit results}

\begin{figure}[h!]
\begin{center}
    \includegraphics[width=0.65\columnwidth]{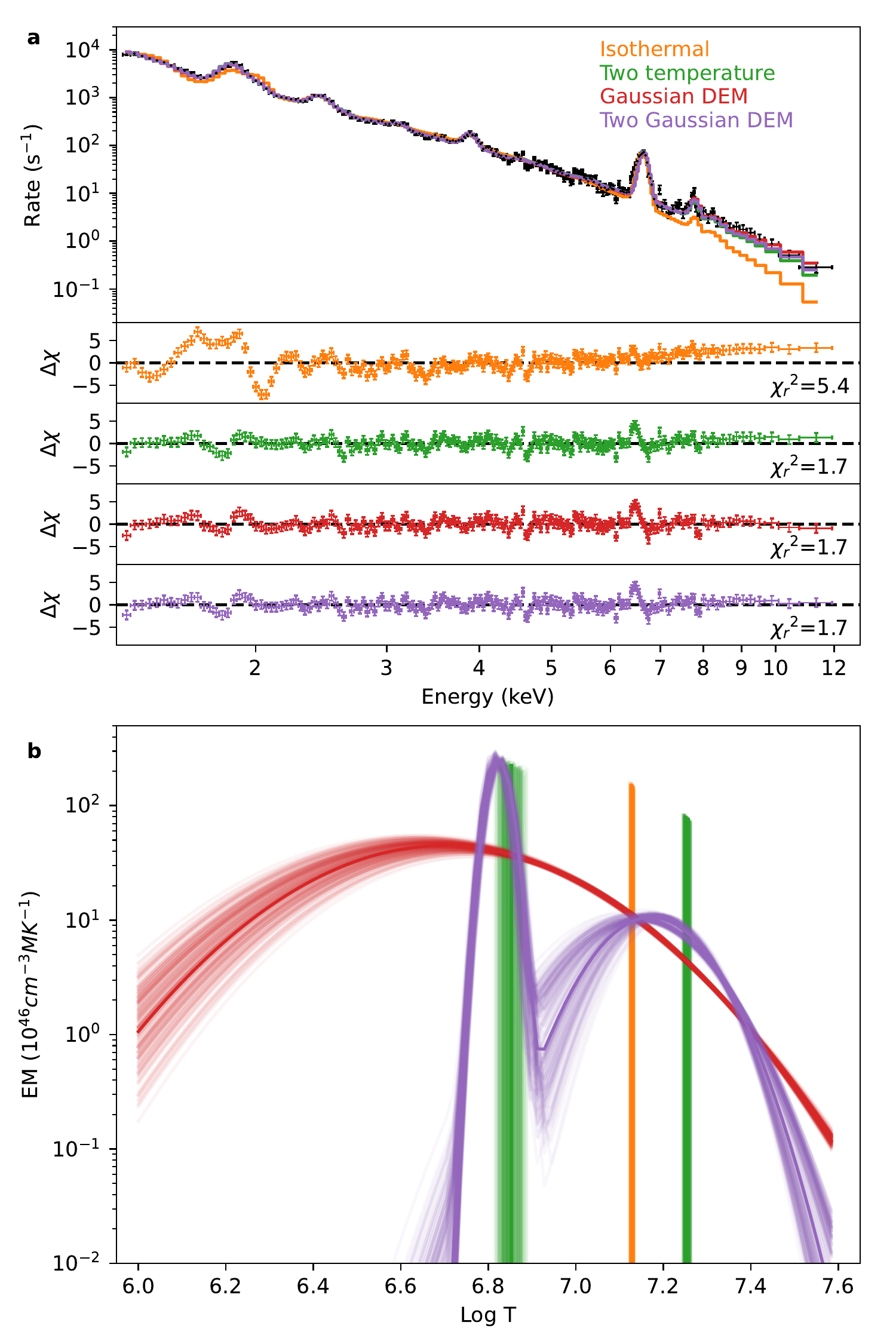}
    \caption{{\bf (a)} XSM spectrum of the impulsive peak of the 08-Sep-2021 flare fitted with different multi-thermal models and 
    isothermal model. The best fit with each model are shown in different colors and the residuals are shown in the lower panels. 
    {\bf (b)} The best fit DEM models from the spectral fit in the panel {\bf a} are shown with the respective colors. 
    Lighter color lines correspond to DEMs from 100 random samples from the MCMC analysis showing the uncertainty 
    of the derived DEM models. 
    \label{bestfitDEM}}
\end{center}
\end{figure}

We carried out the spectral analysis of the flares with these three multi-thermal models 
with the DEM parameters and abundances of elements with prominent lines as 
free parameters. As an example, the results of the fit to the spectrum during the impulsive 
phase of the 08-Sep-2021 flare are shown in Figure~\ref{bestfitDEM}.
The top panel of Figure~\ref{bestfitDEM}a shows the spectrum overplotted with the best fit 
models for each case including the isothermal fit presented in the previous section 
for comparison, whereas the bottom four panels show residuals of the 
fit with different models.    
Figure~\ref{bestfitDEM}b shows the best fit differential emission measure distributions 
obtained in each case. For the two-temperature model and isothermal models, 
vertical lines show the temperature and emission measure. 

From the figure, it can be seen that the multi-thermal models provide a much 
better fit to the observed spectrum as compared to the isothermal model. Residuals 
near the Si line complex and at higher energies are substantially less in the 
case of all multi-thermal models. The reduced chi-squared for the multi-thermal 
models are within the acceptable range as compared to the much higher values for 
the isothermal model. An important point to note is the fact that different multi-thermal 
models predict nearly identical X-ray spectra and hence statistically all 
multi-thermal models provide equally acceptable fits to the observations.

The best fit two-temperature model includes one component at temperatures below 10 MK 
having higher emission measure and another component at higher temperature with a slightly
lower emission measure. The isothermal temperature and emission measure lies in between the 
two temperature components obtained from this fit. The best fit Gaussian DEM has a very broad 
distribution peaking at logT of about 6.8 and having significant emission measure at 
lower and higher temperatures. The double Gaussian DEM model fit resulted in two relatively 
narrower Gaussians having peak temperatures closer to the temperatures obtained from 
the two-T fit to the spectra.  

In order to estimate the uncertainty on the best fit DEM models, we carried out 
a Markov Chain Monte Carlo (MCMC) analysis. MCMC chains were run within PyXSPEC 
with the DEM parameters and abundances as free parameters as in the spectral fitting 
and selected parameters from the chain are written out. DEMs computed from 
a few randomly selected set of parameters from the MCMC chain results are shown as 
thin lines in Figure~\ref{bestfitDEM}b. The spread of these DEMs show 
typical uncertainties in the fitted DEM. 
For the two-temperature models, the lower temperature component is less 
constrained compared to the higher temperature component. In the case 
of the Gaussian model, we note that the DEM at lower temperatures are 
less constrained. This is not a surprising result as XSM spectra have relatively 
lower sensitivity to plasma at those temperatures.

\begin{figure}
\begin{center}
    \includegraphics[width=0.99\textwidth]{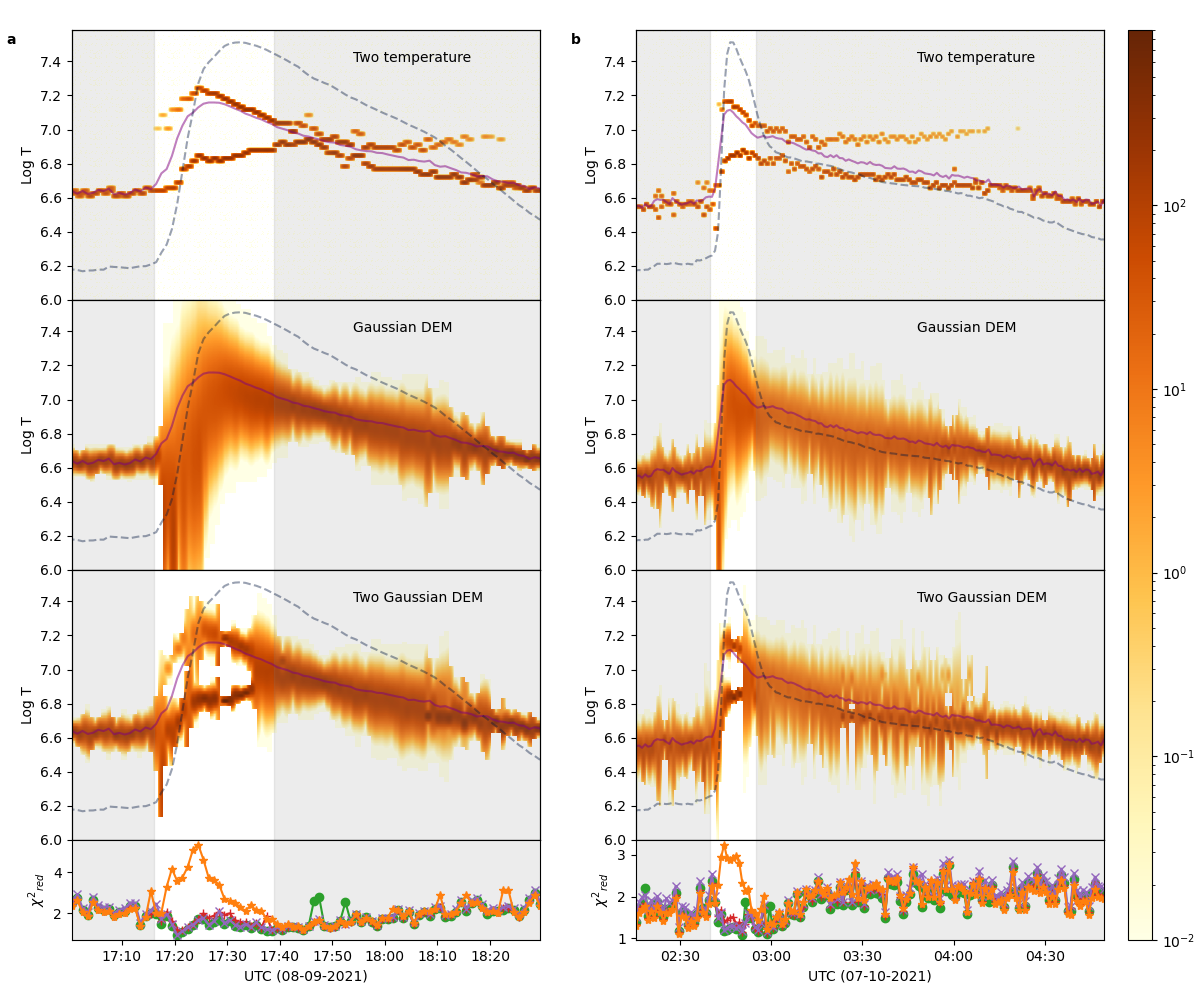}
    \caption{Differential Emission Measure (DEM) evolution during the 08-Sep-2021 flare (a) and 
    the 07-Oct-2021 flare (b) obtained from spectral fitting. The top three panels show DEM with 
    two-temperature, Gaussian, and double Gaussian models, respectively. Color represents the 
    EM in units of $10^{46} cm^{-3} {MK}^{-1}$. The isothermal temperature measurements (purple solid line) and 
    1--15 keV X-ray light curves (dashed line) are also shown for reference.  
    The reduced chi-squared of 
    the fit with two-temperature (green), Gaussian (red), and double Gaussian (purple) are shown 
    in the lower most panels. The reduced chi-squared for the isothermal fit are shown in orange for comparison.
    Intervals when the isothermal and multi-thermal models have a similar fit to the spectra 
    are greyed out and during the remaining period in the impulsive phase, the DEM models provide 
    a much better fit to the spectra in comparison with the isothermal model.
    \label{demEvol}}
\end{center}
\end{figure}

We then fitted the spectrum during each one minute interval with the three models to obtain
the temporal evolution of the flare plasma parameters. As was seen in the previous section, the faintest flare event in the sample is consistent with an isothermal model through out the
duration of the flare. We therefore exclude that event from the multi-thermal analysis. The other
two events were fitted with
each of the three multi-thermal models and the results obtained are shown in
Figure~\ref{demEvol}. Panel a corresponds to the 08-Sep-2021 flare, while panel b  
corresponds to the 07-Oct-2021 flare. For each flare, the three vertical panels show the time evolution 
of the DEM obtained from the two-temperature model, the Gaussian model, and the double Gaussian model as marked 
in the figure. Colors correspond to the value of emission measure at each temperature at respective time 
intervals. 
The isothermal
temperature obtained are overplotted in each case.
The lower most panels
show the reduced chi-square of the fit with each of the multi-thermal models along with 
that from the isothermal fits for comparison.  

From the figure, it can be seen that during the pre-flare period and the decay phase of both the flares 
(marked by grey shade in the figure) 
all three models converge to a narrow range of temperatures close to the isothermal temperature. 
With the two-temperature and double Gaussian models, the temperatures of the two components attain 
the same value (isothermal temperature) during the early and decay phase of the flares. 
This is not very surprising as the spectra during these intervals were consistent with an isothermal model. 
During these periods, there is no notable difference in the reduced chi-square of the fit with different 
multi-thermal models and the isothermal model. 

During the impulsive phase where the isothermal model is clearly not valid, we can see a consistently better 
fit to the spectrum with the multi-thermal model at all time bins similar to the example shown in 
Figure~\ref{bestfitDEM}. Also, the three models have identical quality of fit with the observed spectrum. 
From the evolution of DEM shown in Figure~\ref{demEvol}, we can see that 
the two-temperature fits result in two distinct temperature components during this interval for both 
the flares. A single Gaussian model results in a very broad emission measure distributions during the impulsive 
phase while the double Gaussian model generally shows two narrow Gaussians peaked at distinct temperatures very 
similar to the two-temperature fit. 

The single Gaussian models predict significant EM at lower temperature where soft X-ray 
spectra observed by the XSM has low sensitivity. It is also seen from the MCMC analysis that 
the uncertainties on the DEM at low temperatures are higher. 
While the two-temperature 
DEM is obviously an over simplification, we believe that the two Gaussian DEM is 
most likely the model closer to the true DEM distribution during the impulsive phase 
of the flares.

\subsubsection{Elemental abundances of low FIP elements}

\begin{figure}[h!]
\begin{center}
    \includegraphics[width=0.65\columnwidth]{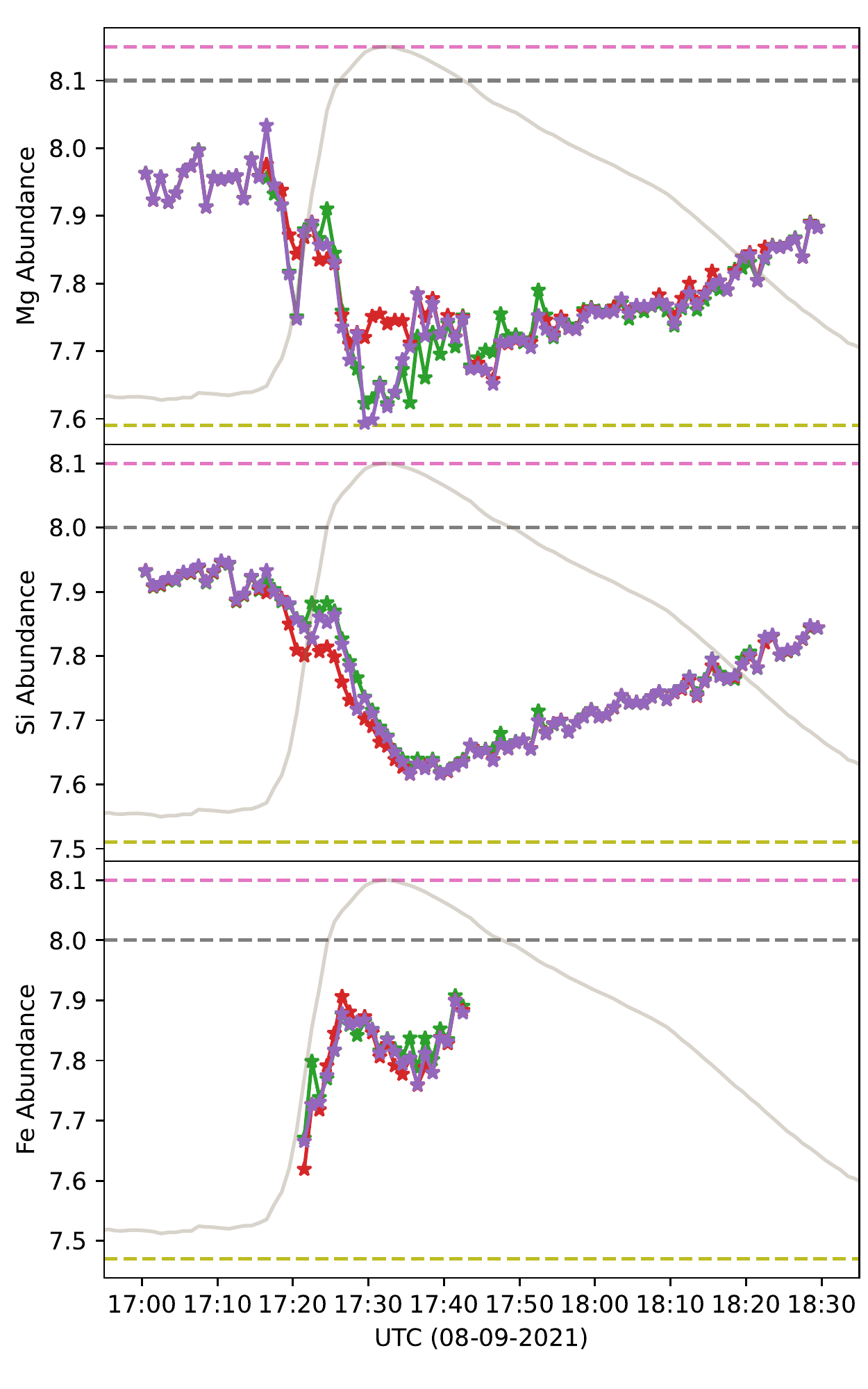}
    \caption{Measured abundances of Mg, Si, and Fe during the 08-Sep-2021 flare obtained 
    from spectral fitting considering three different multi-thermal models: two-temperature (green), Gaussian (red), and 
	double Gaussian (purple), are shown. Coronal abundances from \citet{1992PhyS...46..202F} and \citet{2013A&A...558A..73D} as well as 
    photospheric abundances from \citet{2009ARA&A..47..481A} are shown with purple, grey, and 
    yellow dashed lines, respectively. The X-ray light curve is shown in grey for reference.
    \label{abund}}
\end{center}
\end{figure}

Another interesting result from our time resolved spectral analysis concerns the variation of the elemental abundances. 
In addition to the plasma emission measure distribution parameters, the abundances of elements 
were also free parameters during the fit and we obtain the abundances of various elements during 
the evolution of the flares. Figure~\ref{abund} shows the best fit abundances of the low FIP elements
Mg, Si, and Fe during the course of the 08-Sep-2021 flare. Different colors correspond to the 
abundances obtained from fitting with different multi-thermal models considered in the present 
analysis. 
In the figure, the pink dashed line corresponds to the coronal abundances given by \cite{1992PhyS...46..202F}, the 
grey dashed line 
corresponds to the active region coronal abundances from \citet{2013A&A...558A..73D}, and 
the yellow dashed line corresponds to the photospheric abundances from \cite{2009ARA&A..47..481A}. 

We see that the abundances estimated from three different DEM models match very closely except 
for slight differences seen in the abundances obtained from Gaussian DEM fit during the 
impulsive phase. In any case, the overall trend seen in the variation of abundances during 
the flare is similar with all three models. The abundances of Mg and Si was close to 
the typical coronal abundances before the the flare and they reduce to values 
close to photospheric abundances during the rise phase of the flare. The abundances are 
seen to return to the coronal values during the decay phase of the flare. 
As the Fe abundance 
measurement is available only during the periods near the flare peak, the entire evolution 
is not observed in the case of Fe. 
A similar trend of abundance variations was reported by \citet{2021ApJ...920....4M} for smaller
B-class events observed by the XSM. The present results show that the same effect is seen
during larger flares and that it is seen for Fe abundances too
which was not measurable during the smaller events.

\section{Discussion and Summary}
\label{sec:discussion}

We have presented time-resolved analyses of the soft X-ray spectra of three GOES C-class 
flares using observations with the Chandrayaan-2 XSM. 
While the spectra during the early and decay phases of the flares are consistent 
with an isothermal model, the same is not true for the impulsive phase of the 
higher intensity events (i.e. C5.7 and C8.4 flares).
Spectra were then fitted with three different multi-thermal models where 
DEMs are described with simple parameterized functions. We find that all three
DEM models provide equally good fits to the observations, better than the isothermal model. 
We find that the spectra during the impulsive phase are consistent with the plasma having 
either a very broad temperature distribution or a double peaked distribution.   
The lower temperature part of the broad single Gaussian DEMs are not well constrained with 
XSM observations alone and is possibly adjusted such that the higher temperature 
part fits the spectra well. 
This is not the case with the double peaked DEM 
distributions and thus this is more likely to be closer to the actual plasma parameters.    

The observed characteristics of the DEM evolution during the flares, such as broader DEM at the impulsive phase and narrow DEM during the decay phase, are similar to those reported earlier~\citep{2013ApJ...770..116W, 2014ApJ...788L..31C}.
Double peaked structures of DEM for solar flares observed in this study have also been observed 
using the RESIK X-ray spectrograph~\citep{2006SoSyR..40..125S,2008AdSpR..42..828K}.
If the DEM is indeed double peaked during the impulsive phase, as we infer from the XSM spectra, it is 
likely that the two distinct temperature components correspond 
to the emission from the directly heated plasma in the loop/loop-top and the evaporated chromospheric material filling the flaring loops. 
Hard X-ray imaging observations have shown the presence of coronal loop 
top sources in addition to the chromospheric foot point sources for several 
flares~\citep{2008A&ARv..16..155K}. \citet{2010ApJ...725L.161C} have shown that 
the RHESSI observations of an X-class flare consists of two distinct temperature 
components: one super-hot ($>30$ MK) component located higher in the corona 
and a hot component arising from the foot points. Although the higher temperatures 
in our present observations are not as high as the super-hot component reported in  \citet{2010ApJ...725L.161C}, 
it is not unexpected since the flares we have studied here (C-class) are substantially weaker. 
The fact that the higher temperature component rises faster during the early phase 
of the flare as compared to the lower temperature component further supports the supposition that the 
the former may be associated with the direct heating near the reconnection region.
 
The evolution of abundances of low-FIP elements from coronal to near photospheric and back to 
coronal values are similar to the observations by ~\citet{2021ApJ...920....4M} for B-class 
flares. The reduction of abundances to near photospheric values during the rising phase of 
the flares is consistent with the chromospheric evaporation.
The quick recovery back to coronal abundances favors the involvement of flare induced 
Alfv\'en waves causing fractionation~\citep{2021ApJ...920....4M}.
Unlike Mg and Si abundances that reach very close to photospheric values during the flare peak, 
the Fe abundances are found to be in between photospheric and coronal values at flare peak. 
One possibility is that the abundances of the two distinct temperature components are 
different and that fitting the same abundance parameter for both components 
may have resulted in an intermediate value. 
If the hotter component is indeed arising from coronal loops, it would have coronal 
abundances while the material evaporating from the chromospheric foot points 
would have photospheric abundances. As the hotter component would produce more 
Fe line emission compared to that of Mg or Si (see Figure~\ref{twoTintwoEne}), the abundances measured considering the
same values for both components will effect Fe more and this is reflected in the 
observations. However, this could not be confirmed with the present data as fitting 
with different Fe abundance parameters for the two components did not yield a 
consistent results. 
It is supported by the fact that
previous studies of Fe abundances during flare peaks using the observations of higher temperature lines with RHESSI
by \citet{2012ApJ...748...52P} found abundances closer to coronal values as oppposed to reports 
of close to photospheric abundances from lower temperature lines (for e.g.~\citealp{2013A&A...555A..59D}). 
It is also possible that there are some other factors 
such as non-thermal broadening of Fe lines that are not included in the 
model which could be responsible for the difference in Fe abundance values. Spectral fits shown 
in Figure~\ref{bestfitDEM}a with all multi thermal models show small but visible residuals 
near the Fe line complex. We plan to further investigate this in detail in the future.     

While the XSM spectra can provide some insights into the broader temperature distributions, 
in order to constrain the DEM over a wide range of temperatures and to see where the different 
temperatures might be originating, it would be useful 
to combine observations from multiple instruments sensitive to different 
temperature ranges which also have spatial information.
For example, observations with some channels of SDO AIA are sensitive to 
lower temperatures as compared to the XSM and thus these instruments complement each other.
We plan to take up joint fitting of AIA and XSM observations to have better 
constraints to the lower temperature part of the DEM. \citet{2022ApJ...934..159Z} 
has made comparisons of the model spectrum obtained using AIA DEM with the XSM 
observed spectrum of a B class event to find that they match very well showing that it would be 
possible to carry out such joint fitting.
On other other hand, the DEM associated with higher temperature plasma as well as the non-thermal 
particles can be better constrained with observations in hard X-rays. 
Thus, observations in both soft and hard X-rays of flares will allow us
to constrain the characteristics of both the thermal and non-thermal
particle populations simultaneously, which has implications on improving our 
understanding on the energy partition in flares. 
In this context, such complementary observations for XSM are available 
with the Spectrometer Telescope for Imaging X-rays (STIX, \citealp{2020A&A...642A..15K}) on board the Solar 
Orbiter mission and we plan to use these in the future.

The results from the present work points towards a complex heating 
process of coronal loops during flares. The observation of two distinct temperature components seems to support
the role of direct heating of coronal loop-tops by  magnetic reconnection and secondary heating of post-reconnected loops by chromospheric evaporation. This is also consistent with the evolution of the elemental abundances.
The XSM has been operating for about three years now and is expected to be operational for few more years possibly into the 
next solar maximum. Soft X-ray spectra obtained with the XSM combined with observations in other wavelengths 
provide a unique opportunity for studying the thermal evolution of flares to further our understanding of 
plasma heating processes in solar flares.

\acknowledgments
{
We acknowledge the use of data from the Solar X-ray Monitor (XSM) on board the Chandrayaan-2 mission of the Indian Space Research Organisation (ISRO), archived at the Indian Space Science Data Centre (ISSDC). 
The XSM was developed by Physical Research Laboratory (PRL) with support from various ISRO centers.
We thank various facilities and the technical teams from all contributing institutes 
and Chandrayaan-2 project, mission operations, and ground segment teams for their support.
Research work at PRL is supported by the Department of Space, Govt. of India.
We acknowledge the support from Royal Society through the international exchanges 
grant No. IES\textbackslash R2\textbackslash170199. GDZ HEM  and YR acknowledge support from STFC
(UK) via the consolidated grant to the atomic astrophysics group at DAMTP, 
University of Cambridge (ST/T000481/1).
}
\vspace{5mm}
\facilities{Chandrayaan-2(XSM)}
\software{XSMDAS~\citep{mithun20_soft}, XSPEC~\citep{arnaud96} and PyXSPEC, Python, matplotlib}

\appendix

\section{Effect of temporal averaging}
\label{sec:tavg}

\begin{figure}[h!]
\begin{center}
    \includegraphics[width=0.99\columnwidth]{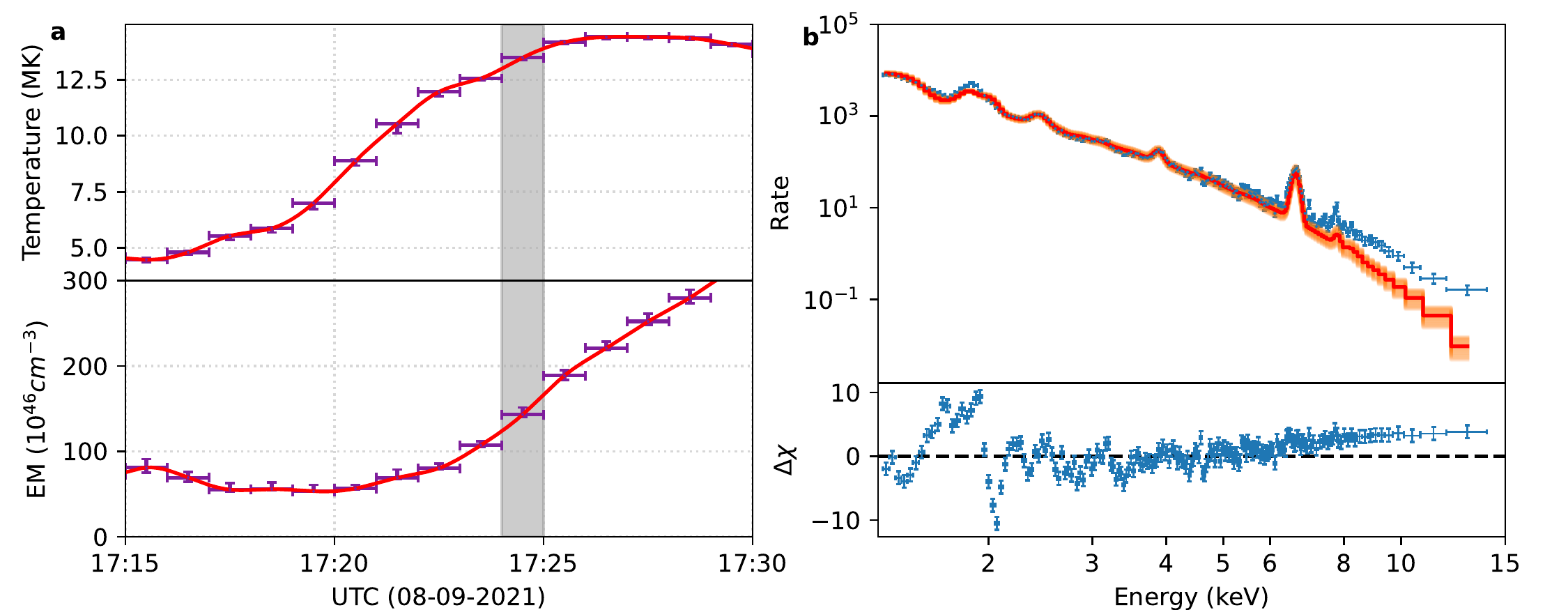}
    \caption{Panel {\bf a} shows temperature and emission measure of the 08-Sep-2021 flare. Red line is interpolated
    values from the measurements at one minute. Spectrum for the grey shaded interval (17:24-17:25) is shown in panel {\bf b} with 
	blue data points.
    Isothermal models corresponding to the interpolated temperature and EM for each one second within the interval
    are shown with light orange lines and the average model considering the variation of temperature and EM
    is shown by solid orange line. Residuals of the observed spectrum with this average model (solid orange line in upper panel) 
    are shown in the lower panel.
    \label{T_EM_interpol}}
\end{center}
\end{figure}

In order to investigate whether the fast evolution of temperature and emission measure during 
the integration period of one minute is the 
reason why the observed spectrum is not consistent with a single temperature, we carried out 
the following analysis. 
As the temperature and emission measure
show smooth variations, we interpolate the fitted parameters from one minute spectra to obtain
the temperature and emission measure for the event on 08-Sep-2022 at every second as shown in Figure~\ref{T_EM_interpol}a.
Then, model spectra for each one minute
interval is obtained by summing the isothermal spectral at each second with the respective
temperature and emission measure. It may be noted that the abundances for each one second 
were set to the best fit value obtained for the one minute interval.
This model is then compared with the observed spectrum as shown
in Figure~\ref{T_EM_interpol}b. We find that this model, which takes into account the
evolution of plasma parameters within the integration time, is also not consistent with
the observed spectrum and has only marginally better goodness of fit compared to
the isothermal fit. We obtain the same result for other time intervals. We also carried out
spectral fitting for shorter integration times and although the statistical uncertainties are
higher, inconsistency of soft X-ray spectra with isothermal models in the impulsive phase
remains apparent. We therefore conclude that the X-ray spectra with the XSM during the impulsive phase
shows the presence of multi-thermal plasma.

\section{Line profiles with XSM and temperature response}
\label{sec:lineprof}
\begin{figure}[h!]
\begin{center}
    \includegraphics[width=0.99\textwidth]{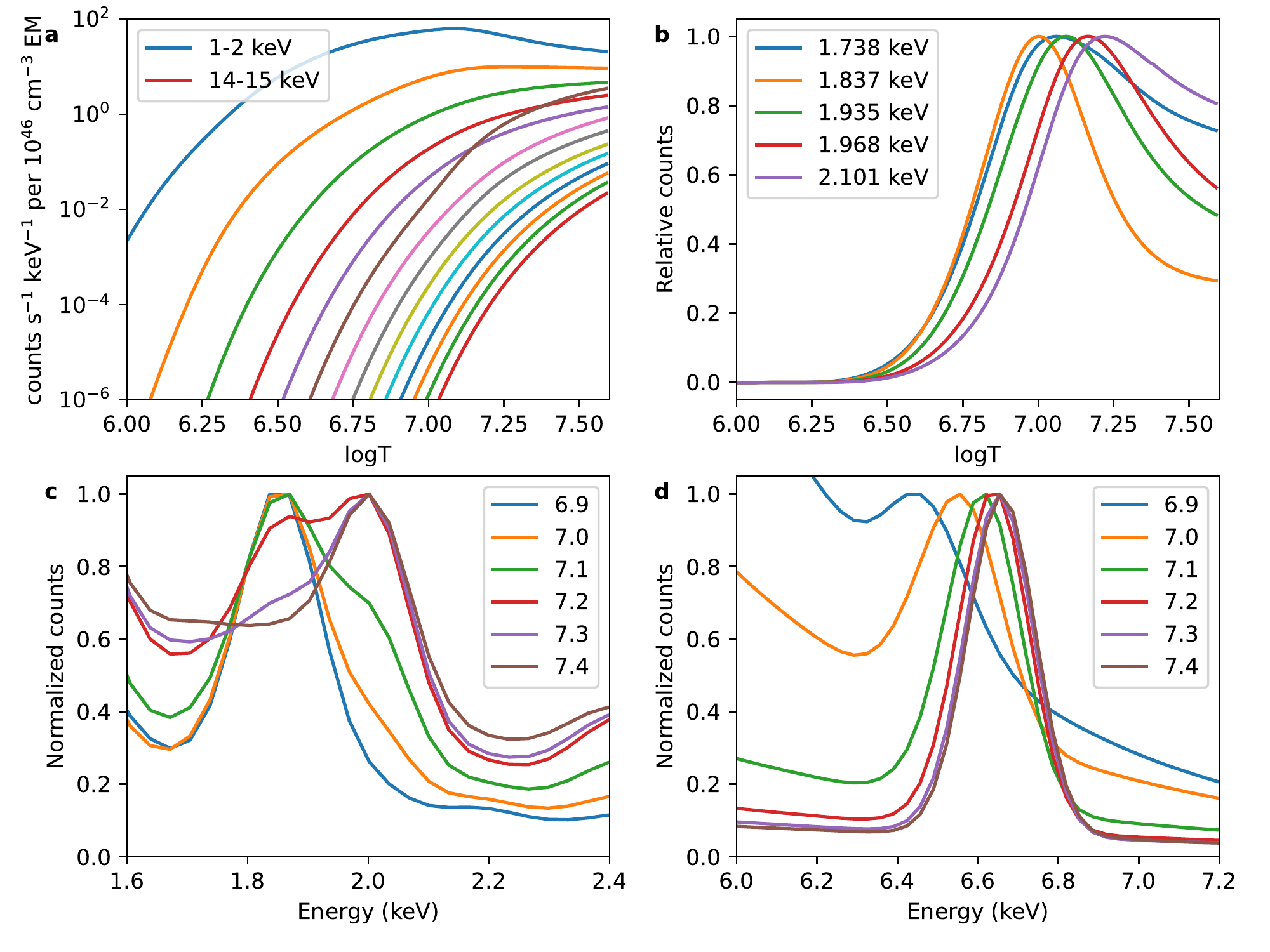}
	\caption{Panel a shows temperature response of the XSM in 1 keV energy bins from 1 to 15 keV 
		in sequence while panel b has normalized temperature responses 
		of selected energy channels near the Si line complex. 
		Simulated XSM spectra near the Si and Fe line complexes at different temperatures are shown in panels 
		c and d, respectively.
    \label{tresp}}
\end{center}
\end{figure}

In order to understand the sensitivity of XSM spectra to different temperatures, we generated temperature
responses of various XSM energy bands. 
Model spectra for a range of temperatures are generated from the
CHIANTI model and then convolved with the XSM response matrix to obtain the counts in different energy
bands arising from plasma at different temperatures. Figure~\ref{tresp}a shows the temperature response of
the XSM energy bands of 1 keV in the entire energy range from 1 - 15 keV. It can be noted that all except two
energy bands have a rather similar shape of the temperature response, as those energy bands are primarily
dominated by continuum emission. The 1-2 keV and 6-7 keV bands include strong line emission from Mg
and Si and Fe, respectively and hence show a slightly different behaviour. The 1-2 keV band has a local
maximum in the temperature response near log T of 7.1. Figure~\ref{tresp}b shows the normalized
temperature response of some of the energy channels of the XSM in the Si line complex, which explains the local
maximum seen in the overall energy range response. It can be seen that different energy channels in the Si
line complex are most sensitive to different temperatures in the range of log T from $\sim$6.9 to $\sim$7.3. This is due
to the different formation temperatures of various lines from different ionization states of Si that are blended
together in the observed line complex and even though the lines are not completely resolved, the spectral
resolution of the XSM allows the disentanglement of the emission from different temperature ranges. We can 
understand this better from panel c of Figure~\ref{tresp} which shows the normalized spectra near the Si line 
complex for different range of temperatures. Similarly, the change in observed line profile for the Fe line complex 
is shown in panel d of Figure~\ref{tresp}, although the differences are less in this case.

We infer that the XSM spectra, especially the Si line complex, shows a distinct response at different 
temperatures allowing the disentanglement of different temperature components in the range of logT 6.9-7.3.  
As the line shape is the most prominent feature to distinguish different temperatures, any multi-thermal 
analysis with XSM spectra use the full spectrum rather than using counts in broader 
energy bins. 

\bibliographystyle{aasjournal}
\bibliography{references_xsm}

\end{document}